\documentclass[%
 reprint,onecolumn,
 amsmath,amssymb,
 aps, showkeys
]{revtex4-2}
\usepackage{amsfonts}
\usepackage{graphicx}% Include figure files
\usepackage{dcolumn}% Align table columns on decimal point
\usepackage{bm}% bold math
\usepackage{stmaryrd}
\usepackage{color}
\usepackage{cancel}
\usepackage{mathtools}
\usepackage{upgreek}
\graphicspath{{./figures/}}

\newcommand{\pfr}[2]{\ensuremath{\frac{\partial #1}{\partial #2}}}
\newcommand{\pfi}[2]{\ensuremath{{\partial #1}/{\partial #2}}} 
\newcommand{\mb}[1]{\mathbf{#1}}
\newcommand\Rey{\mbox{\textit{Re}}}

\begin{document}

\preprint{APS/123-QED}

\title{Axially strained flow in a porous duct of  circular-sector cross-section}% Force line breaks with \\
\author{Prabakaran Rajamanickam} 
 \email{prabakaran.rajamanickam@strath.ac.uk}
\affiliation{Department of Mathematics and Statistics, University of Strathclyde, Glasgow G1 1XQ, UK}

\date{\today}

\begin{abstract}
A canonical problem of axially strained flow in a duct of circular-sector cross-section, with fluid injection through the circular arc, is examined for a range of Reynolds numbers  and sector angles. At small Reynolds numbers, the flow remains symmetric about the mid-plane; however, symmetry is rapidly lost as the Reynolds number exceeds order one, giving rise to asymmetric structures. These flows are characterised by dominant vortices on one side of the duct and secondary vortices on the other when the Reynolds number is sufficiently large. The interaction between interior vortex structures and boundary layers on porous and impermeable walls governs the separation and attachment of the latter.  The axial pressure-gradient coefficient, which also determines the axial strain rate, approaches a constant value at large Reynolds numbers; yet, in contrast to the classical Taylor--Culick result, its value depends sensitively upon duct geometry and flow symmetry. The results underscore the  significance of considering partial-wall injection and asymmetric solutions  in practical applications.
\end{abstract}


\maketitle

%\tableofcontents

\section{Introduction}

When a fluid is injected steadily into a slender porous tube, an axially strained (axisymmetric) flow is established, with the fluid exiting through one or both ends depending on whether the tube is closed at one end or open at both. Away from these ends, the flow admits a self-similar description. For a circular tube, a self-similar solution for laminar viscous flows was described earlier by Yuan and Finkelstein~\cite{yuan1956laminar}, extending the two-dimensional planar problem addressed by Berman~\cite{berman1952laminar}. Taylor~\cite{taylor1956fluid} studied the inviscid limit of injection through porous channels, wedges, ducts and cones, performing  experiments to test his theory.  Culick~\cite{culick1966rotational} applied the self-similar flow to model the combustion chamber of solid propellant rockets, where inward gas flow normal to the chamber wall occurs due to  gasification of the solid propellant, followed by exothermic reactions at the flame adjacent to the wall. 

Taylor~\cite{taylor1956fluid} showed that, for the axisymmetric geometries he considered, the inviscid flow is inherently rotational. In the case of a porous tube, the axial velocity follows the familiar cosine-shaped profile, known as the Taylor--Culick profile. The analogue of Taylor's solution for compressible flows has been studied in~\cite{balakrishnan1992rotational}, revealing several interesting effects. In the Taylor--Culick flow, the vorticity is strictly azimuthal,  and as such there is no vortex stretching by the axial flow. However, Balachander \textit{et al.}~\cite{balachandar2001generation} demonstrated that even slight deviations from axisymmetry can produce axial vorticity, which can then undergo vortex stretching and become significant, particularly near the centerline.

The non-perturbative effects of non-axisymmetry arising from a non-circular cross-section were first analysed by Li{\~n}{\`a}n \textit{et al.}~\cite{linan2004flow}, who laid the groundwork for understanding how such geometries influence the flow. Their  work showed that at large Reynolds numbers,  streamlines originating from the porous wall divide the flow field into regions where some fluid reaches the centerline, while in other places, the fluid swirls and contributes to the formation of complex vortex structures. The study also offered insights into the inviscid limit and the structure of these vortex cores. These initial findings were subsequently extended and refined in later studies~\cite{kurdyumov2006steady,bouyges2017asymptotically}, offering a more detailed description of large Reynolds-number flows in non-axisymmetric geometries.

Besides their applications to solid propellant rockets, axially strained flows in confined geometries have also been analysed in connection with experimental setups designed to study  so-called tubular flames~\cite{wang2006stretch,rajamanickam2021steady}.

\begin{figure}[h!]
\centering
\includegraphics[width=0.45\textwidth]{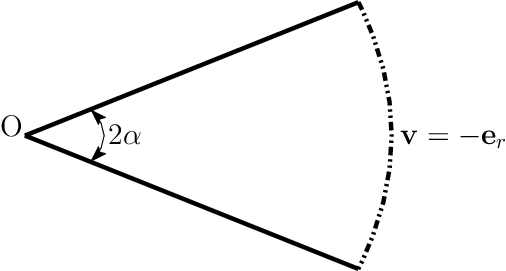}
\caption{Schematic illustration of the circular-sector cross-section with a sector angle $2\alpha$. The curved portion is porous and exhibits the velocity field $\mb v = -\mb e_r$. The length of the each straight segment is unity.} 
\label{fig:wedge}
\end{figure}

The present work aims  to extend the description of such flows to  a new configuration, namely a duct with a circular-sector cross-section. In this geometry, fluid is injected radially inward through the curved, porous part of the sector, while the two straight segments remain impermeable, as illustrated schematically in Fig.~\ref{fig:wedge}. The problem is governed by two parameters, namely the sector angle $2\alpha$ and the Reynolds number $\Rey$, which is defined based on the cylinder radius and the injection speed. Different limiting cases help illustrate the range of flow regimes that can arise. For example, as $\alpha$ approaches zero, the flow is confined to a narrow sector. When $2\alpha=\pi$, the geometry reduces to  a half-circular duct with injection along the curved wall. As $2\alpha$ approaches $2\pi$, the geometry resembles a circular tube with a radial flange. Although this last case might seem analogous to the classical Taylor--Culick flow, the actual flow structure at large Reynolds numbers differs significantly due to flow separation from the radial flange. At large Reynolds numbers, flow separation inevitably occurs near the flat impermeable sidewalls due to the adverse pressure gradient generated by the radially inward flow imposed on the boundary layer. A related phenomenon in a rotating system was discussed heuristically by Moffatt and Duffy~\cite{moffatt1980local}.

\section{Governing equations and boundary conditions}\label{sec:form}

Consider the circular-sector cross-section of unit radial length, bounded by the circular arc, as illustrated schematically in Fig.~\ref{fig:wedge}. We shall use both Cartesian coordinates $(x,y,z)$ and cylindrical coordinates $(r,\theta,z)$ interchangeably. At the porous boundary $r=1$, the velocity field is specified as $(v_x,v_y,v_z)=(-x,-y,0)$. The problem admits a self-similar solution of the form~\cite{linan2004flow}
\begin{equation}
    v_x = u(x,y), \quad v_y=v(x,y), \quad v_z = z w(x,y),\quad 
     p = -\frac{1}{2} (\gamma z)^2 + \Pi(x,y),
\end{equation}
where $\gamma$ is an unknown constant characterizing the axial pressure gradient, or equivalently, the axial strain rate. In developing internal flows, the axial pressure gradient is not determined a priori but is obtained as part of the solution. The corresponding vorticity field is given by
\begin{equation}
    \omega_x = z\pfr{\omega}{y}, \qquad \omega_y= -z\pfr{w}{x}, \qquad \omega_z =\Omega\equiv \pfr{v}{x}-\pfr{u}{y}.
\end{equation}
This self-similar structure applies far from  from ends of the duct, under the assumption that the duct is slender so that its axial length greatly exceeds unity. 

The governing equations for the self-similar solution become
\begin{align}
    \pfr{u}{x}+\pfr{v}{y} &+ w= 0, \label{cont}\\
    u\pfr{u}{x} + v\pfr{u}{y}  &= -\pfr{\Pi}{x} + \frac{1}{\Rey}\nabla^2 u,  \label{xmom}\\
    u\pfr{v}{x} + v\pfr{v}{y}  &= -\pfr{\Pi}{y} + \frac{1}{\Rey}\nabla^2 v, \label{ymom}\\
    u\pfr{w}{x} + v\pfr{w}{y} &+ w^2  = \gamma^2 + \frac{1}{\Rey}\nabla^2 w.  \label{zmom}
\end{align}
where $\nabla^2 = \pfi{^2}{x^2}+\pfi{^2}{y^2}$ is the two-dimensional Laplacian operator. The boundary conditions are given by
\begin{align}
    r=1: & \qquad u = -x, \quad v=-y, \quad w =0,  \label{rbc}\\
    \theta=\pm \alpha: & \qquad u =0, \quad v=0, \quad w=0. \label{tbc}
\end{align}
The unknown constant $\gamma$ is determined by enforcing conservation of mass,
\begin{equation}
    \int_0^1\int_{-\alpha}^{+\alpha} w\, rdrd\theta = 2\alpha. \label{integral}
\end{equation}

An alternative formulation~\cite{kurdyumov2006steady} is possible by replacing the primitive variables $(u,v,w,p)$ with $(\varphi,\psi,w,\Omega)$, where 
\begin{equation}
    u= \pfr{\varphi}{x}+\pfr{\psi}{y}, \qquad v= \pfr{\varphi}{y} -\pfr{\psi}{x}.
\end{equation}
The functions $\psi$ and $\varphi$ should not be interpreted as streamfunction or velocity potential. In this formulation, the governing equations are given by
\begin{align}
    -w = \nabla^2 \varphi,&\qquad \qquad u\pfr{w}{x} + v\pfr{w}{y} + w^2  = \gamma^2 + \frac{1}{\Rey}\nabla^2 w  \label{Alt1}\\
    -\Omega = \nabla^2 \psi, &\qquad \qquad   
    u\pfr{\Omega}{x} + v\pfr{\Omega}{y}   = w\Omega + \frac{1}{\Rey}\nabla^2 \Omega. \label{Alt2}
\end{align}

It is also possible to look for solutions that are symmetric with respect to the mid-plane $\theta=0$ by solving the equations only in the half-domain $\theta \in [0,\alpha]$ and imposing symmetry conditions at $\theta=0$,
\begin{equation}
    \theta=0: \qquad \pfr{u}{y} =v=\pfr{w}{y}=0. \label{symmetric}
\end{equation}
This was the strategy adopted in earlier works~\cite{linan2004flow,kurdyumov2006steady,bouyges2017asymptotically}. While such symmetric solutions are always realized at low Reynolds numbers, they can become restrictive at higher Reynolds numbers. In this study, we shall not assume this symmetry as asymmetric solutions often become the preferred stable configurations at large Reynolds numbers.

\section{Solution for small Reynolds numbers}

As the Reynolds number approaches zero, the nonlinear inertial terms in~\eqref{xmom}-\eqref{zmom} becomes negligible. In this limit, the pressure gradient  must be proportional to $1/\Rey$ in order  to balance the viscous forces. In other words, $\gamma^2\sim 1/\Rey$ and $\Pi \sim 1/\Rey$ as $\Rey\to 0$. At leading order, the governing equations then simplify to
\begin{align}
    \nabla^2 w = - \gamma^2 \Rey, \qquad \pfr{u}{x}+\pfr{v}{y} + w= 0, \quad
     \nabla^2 u = \Rey\pfr{\Pi}{x} , \qquad  \nabla^2 v = \Rey\pfr{\Pi}{y}
\end{align}
subject to the same boundary conditions as the full problem. 

The problem for $w$ is identical to a Poiseuille-flow problem, and the solution for a circular-sector cross-section has been provided in~\cite{moffatt1980local}. We have\footnote{The singularities emerging for $2\alpha=\frac{\pi}{2}$ and $2\alpha=\frac{3\pi}{2}$ are transitional, as shown in~\cite{moffatt1980local}.}
\begin{equation}
    w = \gamma^2\Rey \left\{\frac{r^2}{4}  \left(\frac{\cos 2\theta}{\cos 2\alpha}-1\right) + \frac{2}{\alpha}\sum_{n=0}^\infty \frac{(-1)^{n+1} r^{\lambda_n}}{ \lambda_n (\lambda_n^2-4)} \cos\lambda_n \theta\right\}, 
\end{equation}
where $\lambda_n = (2n+1) \pi/2\alpha$, $n=0,1,2,\dots$. 
Imposing the condition~\eqref{integral}, we find
\begin{align}
      \gamma^2\Rey &= \frac{2\alpha}{\tfrac{1}{16}(\tan2\alpha -2\alpha) -\frac{4}{\alpha}\sum_{n=0}^\infty\frac{1}{\lambda_n^2(\lambda_n^2-4)(\lambda_n+2)}},\\
      &=\frac{8\pi\alpha}{\frac{\alpha\pi}{2}-\hat\gamma-2\ln 2 - \psi\left(\frac{2\alpha}{\pi}+\frac{1}{2}\right) +\frac{\alpha}{\pi}\psi^{(1)}\left(\frac{2\alpha}{\pi}+\frac{1}{2}\right)}, \label{gammasmallre}
\end{align}
where $\hat\gamma$ is the Euler--Mascheroni constant, $\psi$ is the digamma function and $\psi^{(1)}$ is the trigamma function. We have $\gamma^2\Rey \to 12/\alpha^2$ as $\alpha\to 0$, $\gamma^2\Rey = 48\pi^2/(3\pi^2+\pi-48\ln 2)$ when $\alpha=\pi/4$, $\gamma^2\Rey = 8\pi^2/(\pi^2-8)$ when $\alpha=\pi/2$ and $\gamma^2\Rey = 72\pi^2/(9\pi^2-64)$ when $\alpha = \pi$. The last value can be compared with the value $\gamma^2\Rey = 4$, derived for the circular tube (i.e., without the flange at $\theta=\pi$)~\cite{balachandar2001generation}. The parameter $\gamma^2\Rey$ is graphed in Fig.~\ref{fig:gamma} as a function of $\alpha/\pi$.

\begin{figure}[h!]
\centering
\includegraphics[scale=0.5]{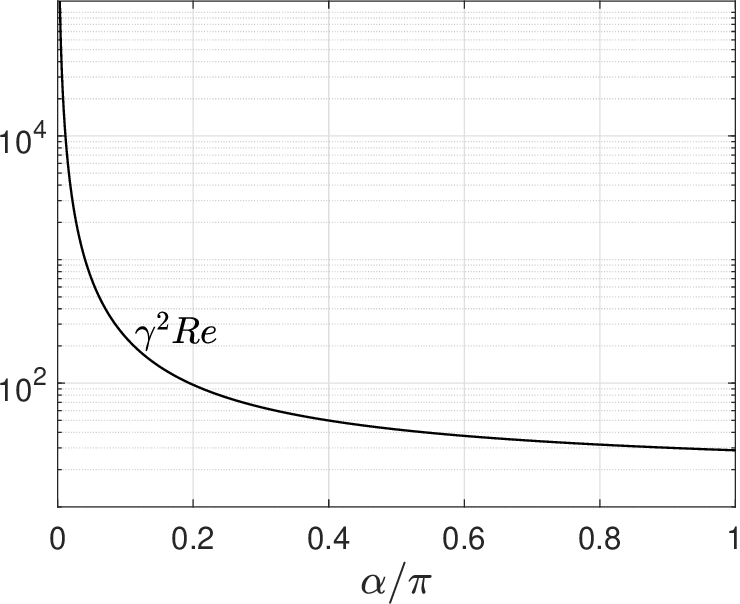}
\caption{Plot of $\gamma^2\Rey$ in the low Reynolds number approximation.} 
\label{fig:smallre}
\end{figure}

Obtaining solutions for the velocity components $u$ and $v$ is not straightforward and generally involve cumbersome calculations. For instance, if we attempt to solve the problem using the alternate formulation mentioned in the previous section, i.e., using~\eqref{Alt1}-\eqref{Alt2} with $\Rey=0$, then the solution for $\varphi$, which satisfies  $\nabla^2\varphi=-w$, is given by  
\begin{align}
    \varphi = -\left(2+\frac{\gamma^2\Rey}{16}\right)\frac{r^2}{2} +\gamma^2\Rey &\left\{\frac{1}{12}\left(\frac{r^2}{2}-\frac{r^4}{4}\right)  \left(\frac{\cos 2\theta}{\cos 2\alpha}-\frac{3}{4}\right)\frac{1}{2\alpha}\sum_{n=0}^\infty \frac{(-1)^{n+1}\cos\lambda_n\theta}{\lambda_n(\lambda_n+1)(\lambda_n-2)}\left(\frac{r^{\lambda_n}}{\lambda_n}-\frac{r^{\lambda_n+2}}{\lambda_n+2}\right) \right\}. 
\end{align}
This solution satisfies the conditions
\begin{align}
     \varphi=0 \quad \text{at} \quad r=0 \quad  \text{and} \quad \pfr{\varphi}{r}=-2 - \frac{\gamma^2\Rey}{16} \quad \text{at} \quad r=1,\\
     -2\alpha \left.\pfr{\varphi}{r}\right|_{r=a} + \int_0^1\frac{1}{r}\left(\left.\pfr{\varphi}{\theta}\right|_{\theta=+\alpha}-\left.\pfr{\varphi}{\theta}\right|_{\theta=-\alpha}\right)dr = 2\alpha.
\end{align}
The last compatibility condition follows from integrating $\nabla^2\varphi=-w$ over the domain and using the condition~\eqref{integral}. Subsequently, one would need to solve the biharmonic equatio
\begin{equation}
    \nabla^4 \psi=0
\end{equation}
subject to appropriate boundary conditions based on the solution for $\varphi$ and consistent with~\eqref{rbc}-\eqref{tbc}. For the purposes of this study, we instead solve the full problem numerically. Moreover, we already have an explicit expression for $\gamma$, which is one of the key quantities of interest in this paper.

\section{Numerical results}

\begin{figure}[h!]
%\centering
\includegraphics[width=0.7\textwidth]{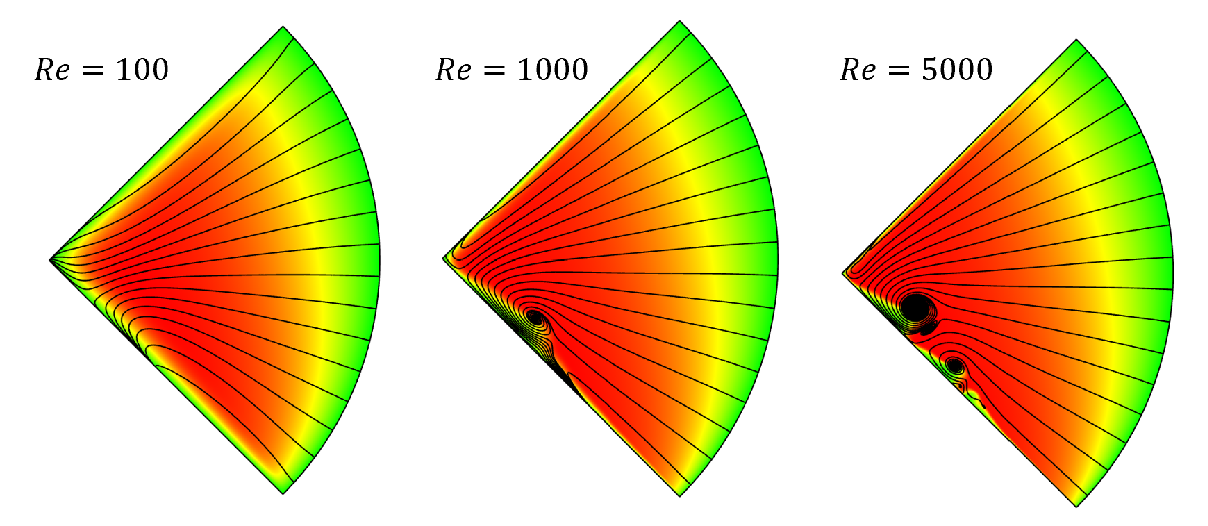}
\includegraphics[width=0.7\textwidth]{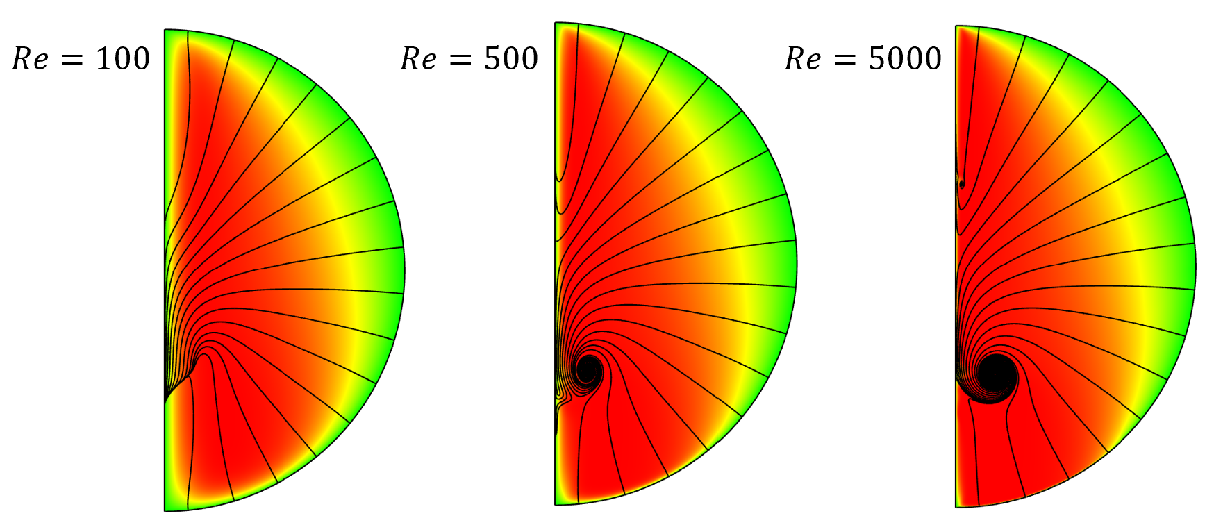}
\includegraphics[width=0.7\textwidth]{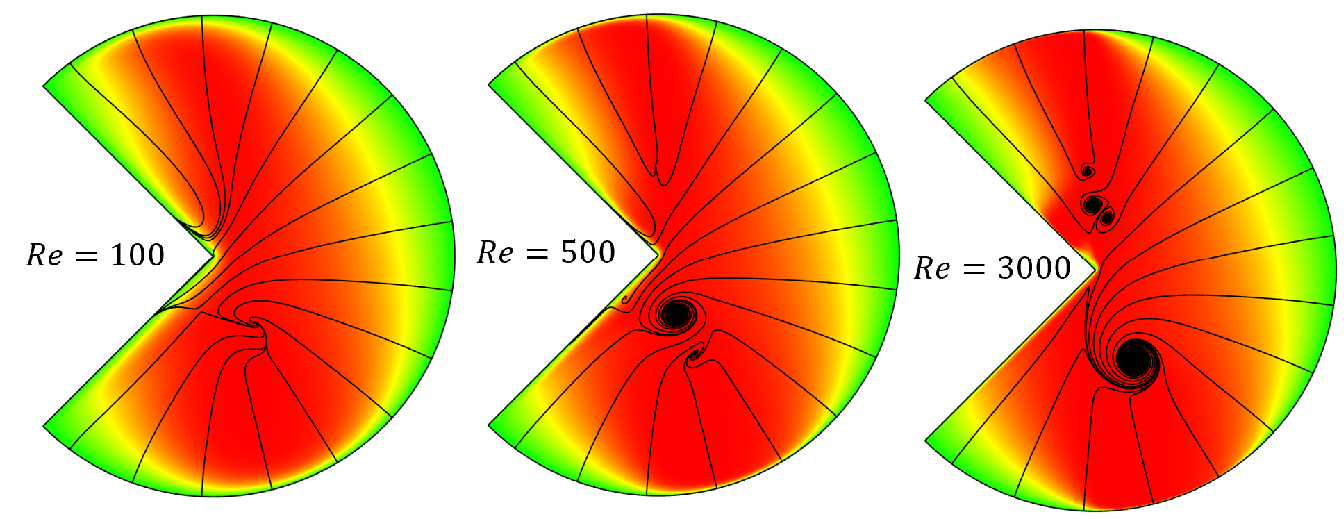}
\includegraphics[width=0.7\textwidth]{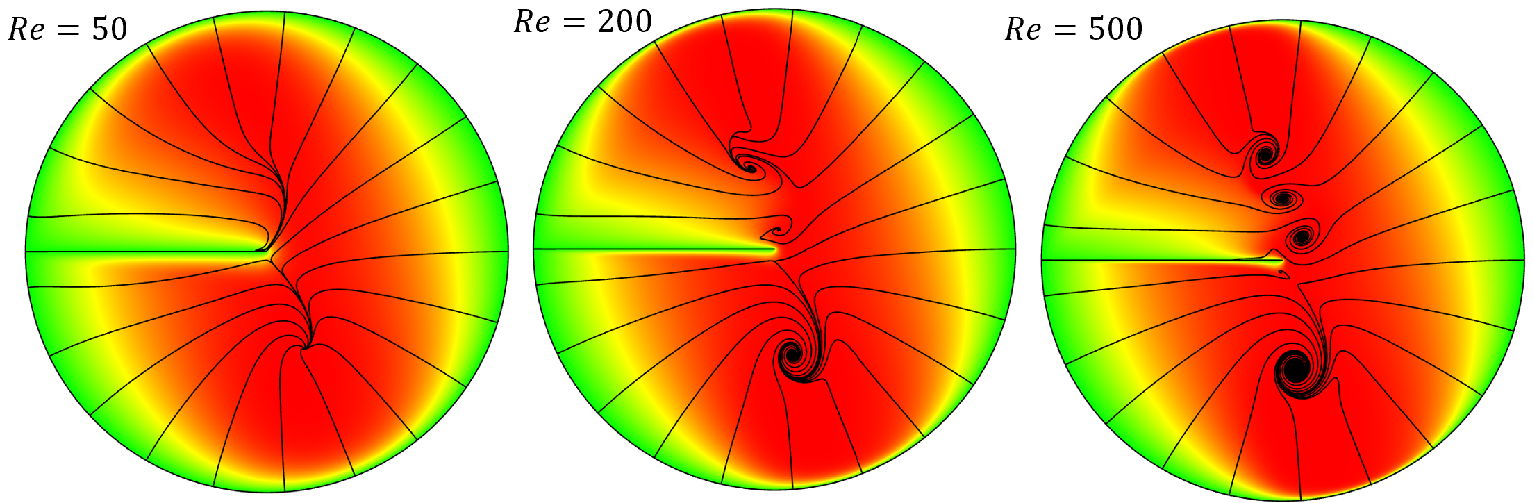}
\caption{The colour contours represent the axial velocity $w$. The projected streamlines corresponds to the velocity components $(u,v)$. First row corresponds to $\alpha=45^\circ$, second row to $\alpha=90^\circ$, third row to $\alpha=135^\circ$ and fourth row to $\alpha\to180^\circ$.} 
\label{fig:stream}
\end{figure}

The solutions for finite values of $\Rey$ are computed numerically by solving~\eqref{cont}-\eqref{integral} using COMSOL Multiphysics software. The computations are performed by parametrically continuing the solutions from small $\Rey$, where a unique solution exists. For the purpose of comparison, we also computed the strictly symmetric solutions by solving the problem in a half-domain as described at the end of Section~\ref{sec:form}. Numerical results for representatives values of $\alpha$ and large values of $\Rey$ are investigated in Fig~\ref{fig:stream}, where the colour map of $w$ is shown along with the projected streamlines formed by the velocity components $(u,v)$. 

From the numerical results, we observed that the solution is symmetric about the mid-plane $\theta=0$ at small values of $\Rey$. However, this symmetry rapidly breaks down as $\Rey$ increases, with the solution becoming noticeably asymmetric already at $\Rey\sim 1$.  In this regime, the axial velocity profile $w$ is slightly stronger on one side of the mid-planer; in our numerical computations, $w$ is typically larger for $\theta<0$. As $\Rey$ is increased further, the asymmetry persists, and flow separation occurs from the impermeable wall on the side where $w$ is stronger. This separation is naturally followed by the formation of a recirculation region adjacent to the wall and, the development of secondary vortex structures as $\Rey$ increases. Such a behaviour is illustrated for the $2\alpha=90^\circ$ case shown in the Fig~\ref{fig:stream}. 

However, for the $2\alpha=180^\circ$ case, one observes the emergence of a strong vortex on one side as $\Rey$ increases, while the boundary layer remains nearly attached along the wall. At $\Rey=5000$, a secondary vortex structure also begins to appear on the opposite side, and it  is expected to strengthen further as $\Rey$ increases.  

The appearance of a dominant vortex structure on one side, accompanied by a sequence of smaller vortices on the other side, occurs at relatively modest values of $\Rey$ as $\alpha$ increases. This behaviour is  evident in the $2\alpha=270^\circ$ and $2\alpha=180^\circ$ cases shown in Fig~\ref{fig:stream}. Interestingly, similar structures were anticipated by Moffatt and Duffy~\cite{moffatt1980local} in a related problem involving flow driven by the rotation of a circular arc~\cite{fraenkel1961corner}. In our configurations, the boundary layer of $w$ remains attached to the impermeable wall on the side where the dominant vortex develops, while it separates on the opposite side. The dominant vortex  and the chain of smaller counter-rotating vortices appear to roughly align along a diagonal line.  Furthermore, $w$ develops a boundary layer on the porous surface on the two ends of the diagonal line, along which vortices are aligned. This finding is noteworthy, since previous studies~\cite{taylor1956fluid,balachandar2001generation,linan2004flow,balakrishnan1992rotational}  implicitly assumed that boundary layers of $w$ would always  be blown off by the incoming flow. In contrast, our results show that the interior vortex patterns and their induced velocities can stabilise and sustain the boundary layer of $w$ near porous walls.

Additionally, Fig~\ref{fig:stream} shows that the orientation of the dominant vortex is flipped in the $2\alpha=180^\circ$ case. Further numerical computations showed the flip in orientation occurs at a critical opening angle of  $2\alpha\approx 316^\circ$.

\begin{figure}[h!]
%\centering
\includegraphics[width=0.7\textwidth]{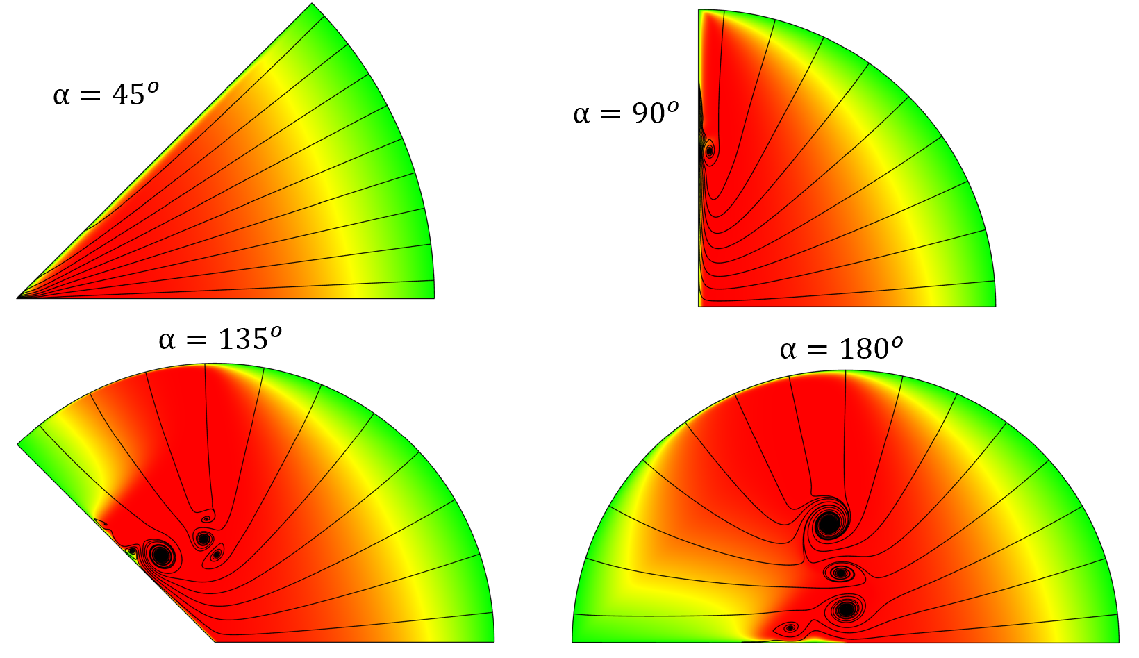}
\caption{The colour contours represent the axial velocity $w$. The projected streamlines corresponds to the velocity components $(u,v)$. The first three figures pertains to $\Rey=2500$, whereas the last case belongs to $\Rey=1000$.} 
\label{fig:symmetric}
\end{figure}

Fig.~\ref{fig:symmetric} presents a few results illustrating the symmetric solutions. In this case, the onset of flow separation is delayed to higher values of $\Rey$. For instance, with $2\alpha=90^\circ$, no discernible flow separation is observed even at $\Rey=2500$, highlighting the importance of asymmetric solutions. Here too, we observe the formation of a dominant leading vortex, followed by a sequence of weaker secondary vortices.

\begin{figure}[h!]
\centering
\includegraphics[scale=0.6]{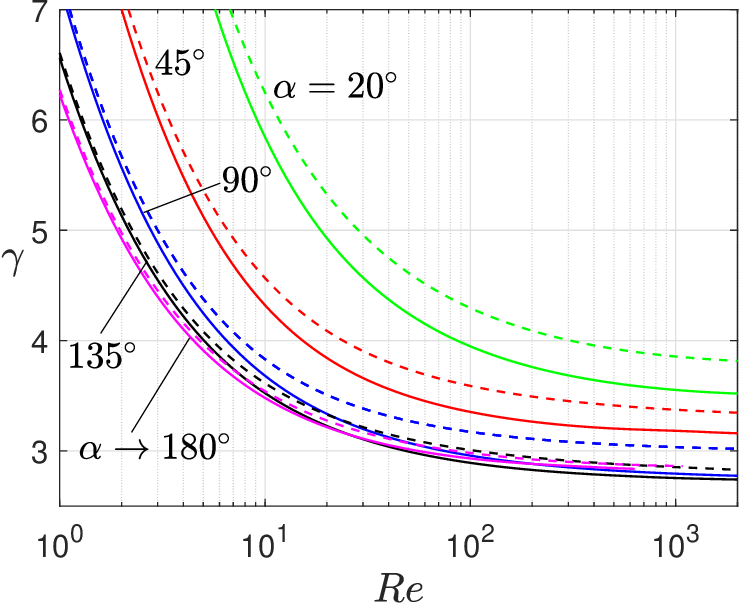}
\caption{Computed values of the parameter $\gamma$ for different sector angles. The solid lines correspond to asymmetric solutions and the dashed lines to symmetric solutions.} 
\label{fig:gamma}
\end{figure}

Finally, the axial pressure gradient coefficient $\gamma$ is plotted in Fig.~\ref{fig:gamma} as a function of $\Rey$ for different values of $\alpha$. The solid lines correspond to asymmetric solutions and the dashed lines to symmetric solutions.
While the parameter $\gamma$ becomes singular at low $\Rey$ as predicted earlier, it approaches a constant value as $\Rey \to \infty$. Importantly, these differ depending on the sector angle and the symmetry of the solution. For large Reynolds numbers, the values can be compared with the corresponding value  for the Taylor--Culick profile 
\begin{equation}
    v_r = -\frac{1}{r}\sin\frac{\pi r^2}{2}, \qquad v_\theta=0, \qquad w=\pi \cos\frac{\pi r^2}{2}, \qquad \gamma = \pi. 
\end{equation}
In fact, Kurdyumov~\cite{kurdyumov2006steady} conjectured that the coefficient $\gamma$ approaches the Taylor--Culick profile as $\Rey\to \infty$ independent of the duct cross-section, provided that injection occurs over the entire cross-section. In our case, however, injection is restricted to only a portion of the duct geometry. The numerical results thus suggest that, under such conditions, $\gamma$ doe not converge to a universe constant at large Reynolds numbers, but instead depends sensitively on both the duct cross-section and the symmetry properties of the flow.

The large Reynolds number limit is challenging to analyse due to the occurrence of flow separation from impermeable walls. However, in regions where the flow separation event does not play a significant role (for example, on the dominant-vortex side of the domain, where the boundary layer remains attached almost everywhere), only can examine the inviscid limit ($\Rey\to \infty$). In this limit, as shown in~\cite{linan2004flow}, the governing equations reduce to
\begin{align}
    \left(u\pfr{}{x} + v \pfr{}{y}\right)p' =0, &\qquad \left(u\pfr{}{x} + v \pfr{}{y}\right)w =\gamma^2-w^2,\\
    \left(u\pfr{}{x} + v \pfr{}{y}\right)\Omega = w\Omega, &\qquad \left(u\pfr{}{x} + v \pfr{}{y}\right)[\Omega\sqrt{\gamma^2-w^2}]=0,
\end{align}
where $p'=\Pi + \tfrac{1}{2}(u^2+v^2)$ is the stagnation pressure. Integrating these along a streamline, starting from the porous surface $r=1$ where the streamlines originate, we obtain
\begin{equation}
    p'=p'(0), \qquad w = \gamma\tanh(\gamma s), \qquad \Omega = \Omega(0)\cosh(\gamma s), \qquad \Omega\sqrt{\gamma^2-w^2} = \Omega(0) \gamma.
\end{equation}
where $s$ is the arclength of the streamline measured from the porous surface. As the streamlines spiral into the vortex, $s$ becomes very large. In the limit $s\to \infty$, we have 
\begin{equation}
    w\to \gamma -2\gamma e^{-2\gamma s}+\cdots \quad \text{and} \quad \Omega\to \tfrac{1}{2}\Omega(0)e^{\gamma s} + \cdots,
\end{equation}
i.,e $w$ reaches its maximum value $\gamma$, while the vorticity grows without bound. The resulting singular vortex is regularised by an inner viscous core, which corresponds to a Burgers'-type vortex structure due to the presence of axial strain in the vortex structure.

\section{Closing remarks}

A canonical problem of axially strained flow in a duct with circular-sector cross-section is investigated. Such flows have potential practical relevance in solid-propellant rockets, tubular flame configurations, and other specialised manufacturing or laboratory setups involving porous surfaces. For instance, Taylor's classical work~\cite{taylor1956fluid}  was motivated by an application arising in the paper-making process. 

The main conclusion to be drawn from this study is that asymmetric structures emerge at moderately large Reynolds numbers, giving rise to complex vortex patterns that are fundamentally different from the class of strictly symmetric solutions. Flow symmetry is inherently unstable beyond a moderate $\Rey$, leading to rich vortex structures that cannot be captured by traditional symmetric analyses, considered in the literature.  The study further  demonstrates that interactions between interior vortex structures and boundary layers on porous surfaces can significantly modify the flow, with boundary-layer separation occurring selectively. Moreover, unlike classical results for full cross-section injection, the axial pressure gradient coefficient $\gamma$ (i.e., the axial strain rate) depends sensitively on duct geometry and flow symmetry, highlighting the importance of partial-wall injection in practical designs.

Beyond the porous-wall problem, the self-similar description of axially strained flow also appears in a related problem where duct walls undergo linear stretching along the axial direction. The stretching-tube problem was first studied by Brady and Acrivos~\cite{brady1981steady}, who demonstrated the existence of multiple solutions. Unlike the present problem, in that case the coefficient $\gamma^2$ can take both positive and negative values. It would be worthwhile to investigate the existence of symmetric and asymmetric solutions for such problems in ducts with non-circular cross-sections in future studies.

\bibliography{references}

\end{document}